\documentclass{article}
\usepackage{graphicx} 
\usepackage{biblatex}
\usepackage{amsthm, amsmath, amsfonts}

\theoremstyle{definition}
\newtheorem{definition}{Definition}[section]

\newtheorem{theorem}{Theorem}[section]

\title{Consensus, Inconsistency, Emergence:\\
\large what's paraconsistency got to do with it?}
\author{
Gabriel Rocha\\
\small Centre for Logic, Epistemology and the History of Science\\
\small Universidade Estadual de Campinas (Unicamp) \\
\small \texttt{gabrielrocha.comp [at] gmail.com}}
\date{July 2025}

\begin{document}

\maketitle

\begin{abstract}
    The consensus problem, briefly stated, consists of having processes in an asynchronous distributed system agree on a value. It is widely known that the consensus problem does not have a deterministic solution that ensures both termination and consistency, if there is at least one faulty process in the system. This result, known as the FLP impossibility theorem \cite{10.1145/3149.214121}, led to several generalizations and developments in theoretical distributed computing. This paper argues that the FLP impossibility theorem holds even under a generalized definition of computation through oracles. Furthermore, using a theoretical machinery from complex systems, this paper also posits that inconsistency may be an emergent feature of consensus over distributed systems by examining how a system transitions phases. Under the same complex systems framework, this paper examines paraconsistent logics, arguing that while inconsistency is not an emergent feature for these logics, triviality may be. Lastly, some attention is given to the possibility of developing consensus algorithms capable of paraconsistent reasoning.
\end{abstract}

\section{Introduction}

The consensus problem is a fundamental distributed computing problem concerning processes that have an initial value, communicate among each other, and must agree on a value. A solution to the consensus problem usually consists of a distributed algorithm called a protocol, In general, one strives to construct a protocol with certain properties relating to fault tolerance. Specifically, it is desirable for a protocol to guarantee three properties:

\begin{itemize}
    \item \textbf{Termination}: every process eventually decides on a value.
    \item \textbf{Consistency}: if two processes $u$ and $v$ decide on values $\nu$ and $\nu'$ respectively, then $\nu = \nu'$.
    \item \textbf{Non-triviality}: if a process $u$ decides on $\nu$, then $\nu$ is one of the initial values.
\end{itemize}

If all processes are fault-free, that is, they always decide on a value and communicate reliably, then one can always construct a deterministic protocol with all three properties if the initial values are totally orderable. However, if even a single process is faulty, such a protocol cannot exist. This result is due to the widely known FLP impossibility theorem \cite{10.1145/3149.214121}: that is, it is not possible to construct a deterministic protocol that ensures both termination and consistency.

The FLP impossibility theorem spawned a vast literature on theoretical distributed computing, its impact yet even further underlined by winning the Dijkstra prize in 2001. This impossibility result led to different models of distributed systems, ranging from fully synchronous to fully asynchronous, with resource bounds and randomization \cite{fich2003hundreds,ASPNES1993414}.
Furthermore, it also led to the similar CAP theorem \cite{6133253,798396} in the context of database theory, with ramifications in several real world domains.

Since deterministic consensus protocols can guarantee non-triviality but not both consistency and termination, one may say that \textit{computable}\footnote{As in, decidable by a Turing Machine.} deterministic consensus is inconsistent, yet not trivial. As so it happens, the lemma ``inconsistent yet non-trivial'' is closely related to da Costa's Principle of Tolerance \cite{daCosta1958} (later re-branded as Principle of Non-Triviality in \cite{carnielli2002taxonomy}) --- briefly stated, any mathematical theory is admissible as long as it is not trivial. da Costa pioneered the groundwork for what now are called paraconsistent logics and, in such logics, inconsistent or contradictory statements do not lead to reasoning collapse.

Furthermore, the property of consistency in a consensus protocol is not directly related to consistency in the processes of a distributed system. That is, consistency or lack thereof arises only when one observes the outcome of the execution of a protocol in the system as a whole. If one observes a single process in the system, there is no inconsistency in the value it is deciding. At first glance, it may seem reasonable to characterize inconsistency in distributed systems as an instance of emergent behavior. However, the problem of precisely defining `emergence' is itself a notoriously difficult endeavor \cite{abrahao2022emergence}, with several conflicting definitions in the literature.

This paper will briefly attempt to connect the concepts of consensus, inconsistency, paraconsistency and emergence in the context of asynchronous distributed systems. The overall goal is not to establish a particular result strongly relating these concepts, rather, it is to build bridges for further exploration.

\section{Background}

Drawing from \cite{charronbost:hal-03721465}, this section will present the computing model that formalizes asynchronous distributed systems. Some of the definitions will be suppressed for brevity.

\begin{definition}[Distributed System]
    A \textit{distributed system} is a graph $G = \langle V, E \rangle$ such that each node $p \in \mathcal{V}$ is a process loosely understood to be a Turing Machine, and each edge represents a fully reliable communication channel between processes.
    
    A message in a distributed system is an element of a set $\mathcal{M}$. Each process is able to \textit{send} and \textit{receive} messages to processes according to the edges in the graph, upon receiving a message a process may transition its state according to some computation.
\end{definition}

\begin{definition}[Configuration]
    A \textit{configuration} of a distributed system is a tuple $C = \langle \sigma_{p_1}, \sigma_{p_2}, \ldots, \sigma_{p_n}, M \rangle$ where $\sigma_{p_1}$ is the state of process $p_1 \in V$ and $M$ is the state of messages in the communication channels.
    An \textit{initial configuration} is a configuration such that there are no messages in the communication channels.
    The sets of all configurations and all initial configurations are denoted $\mathcal{C}$ and $\mathcal{I}$, respectively.
\end{definition}

\begin{definition}[Algorithm Execution]
    An \textit{execution} of a distributed algorithm $\mathcal{A}$ corresponding to a transition system $\langle \mathcal{C}, \mathcal{I}, \vdash \rangle$ is a finite or infinite sequence of configurations $\langle C_0, C_1, \ldots, C_t, \ldots \rangle$ such that:
    \begin{enumerate}
        \item every $C_t$ is a configuration of $\mathcal{A}$, i.e. $C_t \in \mathcal{C}$
        \item $C_0$ is an initial of $\mathcal{A}$, i.e. $C_t \in \mathcal{C}$
        \item for every index $t$, $C_t \vdash C_{t+1}$
    \end{enumerate}
\end{definition}

\begin{definition}[Consensus Decision]
    Let $G$ be a distributed system, $\mathcal{V}$ be a set of values such that $\bot \not \in \mathcal{V}$. Each process $p$ of $G$ has an initial value\footnote{One may consider the `initial value' to be the initial value of a tape, and the output variable to be another tape in a multi-tape Turing Machine.} $\mu_p \in \mathcal{V}$ and an output variable $y_p \in \mathcal{V} \cup \{\bot\}$, initially set to $\bot$.

    When, during the execution of an algorithm $\mathcal{A}$, a process $p$ assigns some value $v \in \mathcal{V}$ to $y_p$, it is said that $p$ decides $v$.
\end{definition}

In the context of the consensus problem, a process $p$ is \textit{faulty} if, during the infinite execution of a distributed algorithm $\mathcal{A}$, it crashes. Note that `crashing' may refer to the process entering an infinite loop or simply not changing state.

\begin{definition}[Admissible Execution]
    A complete execution $\alpha$ of an algorithm $\mathcal{A}$ is admissible if
    \begin{enumerate}
        \item every message sent in $\alpha$ to a correct process is eventually received in $\alpha$;
        \item there is at most one faulty agent in $\alpha$
    \end{enumerate}
\end{definition}

Formally speaking, the FLP impossibility theorem refers to admissible executions of consensus algorithms.

\begin{theorem}[FLP]
    There is no algorithm $\mathcal{A}$ such that every of its admissible executions simultaneously satisfy termination, consistency, and non-triviality.
\end{theorem}

With respect to logic, the concepts of consistency and triviality are generally defined from the syntactical perspective.

\begin{definition}[Logical Consistency and Triviality]
    A logic $\mathbf{L}$, with language $L$, is said to be \textit{consistent} if $ \vdash_\mathbf{L} A$ if and only if $\not \vdash_\mathbf{L} \lnot A$, where $\lnot$ is a negation operator. A set $\Gamma \subseteq L$ is inconsistent if $\{A, \lnot A\} \subseteq \Gamma$.

    Furthermore, $\mathbf{L}$ is called \textit{trivial} if $\vdash_\mathbf{L} A$ for any $A \in L$. If $\Gamma \vdash_\mathbf{L} A$ for any $A \in L$, $\Gamma$ \textit{trivializes} $\mathbf{L}$.
\end{definition}

\begin{definition}[Logical Paraconsistency]
    A logic $\mathbf{L}$, with language $L$, is called \textit{paraconsistent} if there is an inconsistent $\Gamma \in L$ which does not trivialize $\mathbf{L}$.
\end{definition}

Recalling the previous section, it is most difficult to clearly define what constitutes emergence. As both distributed systems and logics are objects of formal knowledge, it is reasonable to set the working definition of emergence to be that which is defined over phase transitions \cite{abrahao_2025_15848688}.

\begin{definition}[Features and Phases]
    Let $\mathcal{F}_1$ be a feature of a system $\mathcal{S}$, not necessarily a distributed system. $\mathcal{F}_1$ is \textit{measured} by a function $g : X_1 \rightarrow Y_1$.
    Let $\mathcal{P}_1, \mathcal{P}_2, \ldots$ be the phases through which a system undergoes. $P_i$ is \textit{measured} by a function $f_i$ on domain $Y_1$.
\end{definition}

\begin{definition}[Phase Transition]
    A system $\mathcal{S}$ undergoes a \textit{phase transition} from $\mathcal{P}_i$ to $\mathcal{P}_{i+1}$ by increasing a feature $\mathcal{F}_1$ if there is an $x_{i+1} \in X_1$ such that:

    $$ \forall x \in (x_i, x_{i+1} ]~\left(~ f_i(g(x)) \leq g(x) \leq f_{i+1}(g(x)) ~\right) $$
\end{definition}

Emergence in a system is then defined as a phase transition that can be intuitively ``postponed'' by adding some components to the system, such as additional knowledge, that may alter the measurement function, but the phase transition is guaranteed to occur at some point. Note that this definition differs from the one presented in \cite{abrahao_2025_15848688}.

\begin{definition}[Emergence]
    \label{def:emergence}
    A system $\mathcal{S}$ exhibits \textit{emergent behavior} over a feature $\mathcal{F}_1$ measured by $g$ if:
    \begin{enumerate}
        \item it undergoes at least one phase transition from $\mathcal{P}_i$ to $\mathcal{P}_{i+1}$, and
        \item updating $g$ to a new $g'$ such that $g'(x_i) > f_i(g(x_i))$ eventually leads to the \textit{same} phase transition $\mathcal{P}_i$ to $\mathcal{P}_{i+1}$.
    \end{enumerate}
\end{definition}

\section{Connections}

As a first step, it is warranted to make a distinction between inconsistency in distributed systems and in logic. In distributed systems, inconsistency refers to a system-wide feature. That is, the execution of a consensus algorithm $\mathcal{A}$ over a system $G = \langle V, E\rangle$ \textit{causes} inconsistency in the system if a transition $C_t \vdash C_{t+1}$ results in disagreement of decisions across the processes of $G$. However, the processes themselves are not capable of causing inconsistency. Consider a single process $p$ of $G$, when it decides, it may have some information about the state of $G$ (that is, whether other processes have decided), but it cannot cause decision inconsistency in isolation.

\subsection{Emergence and Consistency}

Let termination, consistency and non-triviality be features $\mathcal{F}_t$, $\mathcal{F}_c$ and $\mathcal{F}_n$ of a distributed system in the execution of a consensus algorithm $\mathcal{A}$. Suppose that $\mathcal{F}_t$, $\mathcal{F}_c$ and $\mathcal{F}_n$ are measured by functions $g_t, g_c, g_n :\mathcal{C}\rightarrow \{T,F\}$, respectively. Suppose also that the number of faulty processes in $G$ is also a feature $\mathcal{F}_\infty$ measured by $g_\infty : \mathbb{G} \rightarrow \mathbb{N}$, where $\mathbb{G}$ is the set of all distributed systems. Phase transitions then may arise under two different perspectives:

\begin{enumerate}
    \item By the FLP impossibility theorem, there is a phase transition when the number of faulty processes grows from 0 to 1, i.e. $g_\infty(G) = 1$. That is, when the number of faulty processes is 0, then $g_t(C) =g_c(C) = g_n(C) = T$ for all configurations $C$, and when the number of faulty processes is 1, that is no longer the case, since there is a configuration $C'$ such that $g_t(C') = T$, yet $g_c(C') = F$ or vice versa. In this case, the phase transition is known `in principle'.
    \item Once again by the FLP impossibility theorem, forcing termination forces inconsistency in the presence of one faulty process. Therefore, if $g_t(C) = T$ for all configurations, there are configurations $C_t$ and $C_{t+1}$ such that $g_c(C_t) = T$, $g_c(C_{t+1}) = F$, and $C_t \vdash C_{t+1}$. This phase transition is, in general, not known `in principle'.
\end{enumerate}

A question then arises: if inconsistency is the result of a phase transition in deterministic consensus, is it emergent behavior? As there are two perspectives on the phase transitions, two answers are warranted.

Starting with the second perspective, it is somewhat straightforward to see that a phase transition occurs in a postponable setting. One can modify $\mathcal{A}$ to define a new $\mathcal{A}'$ such that a new configuration is introduced between $C_t$ and $C_{t+1}$ simply by defining that all processes should send dummy messages at $C_t$, that is, messages that do not interfere with the consensus decision. So in $\mathcal{A}'$ there is a new configuration $C'$  such that $C_t \vdash C'$ and $C' \vdash C_{t+1}$, but $\mathcal{F}_c(C_t) = \mathcal{F}_c(C')$. What this entails is that one can extend the execution of a consensus algorithm indefinitely to postpone the inconsistency. However, by the FLP impossibility theorem, inconsistency is unavoidable if the algorithm terminates. Hence, by definition \ref{def:emergence} of emergence, one may say that inconsistency is an emergent behavior in the context of a consensus algorithm which is guaranteed to halt.

Turning to the more delicate first perspective, one may posit naively that the answer is no --- the phase transition which happens when moving from zero to one faulty process cannot be delayed, and, as a consequence, cannot it cannot be classified as emergent behavior. However, consider the introduction of fault detection oracles\footnote{Oracles are well-known objects used to study barrier problems in complexity theory, such as in \cite{10.1109/CCC.2006.32, 10.1145/1806689.1806711}. However, the term `oracle' here slightly differs from the usage in the literature.}. Suppose that, given a system $G = \langle V, E \rangle$, there is one process $u$ known to be faulty. A process $O_u$ is an oracle for $u$ if, when a process $p$ sends a message to $O_u$, $O_u$ sends a message back stating whether $u$ crashed or not.

Given an algorithm $\mathcal{A}$ for consensus, one may modify it such that each process first sends a message to $O_u$ prior to sending any message process and prior to changing states. This effectively negates the faultiness of $u$, since all processes can reliably query whether or not to consider $u$ in the consensus process. Therefore, with one oracle and one faulty process, it is possible to recover consistency, termination and non-triviality. Now, if another process $v$ is known to be faulty, one may repeat the same process with another oracle $O_v$ and thus the three properties are again recovered, yet the number of faulty processes and oracles is now 2. Notice that this procedure may be extended for a number of faulty processes $|V| - 1$, where $|V|$ is the original number of processes in the system. Furthermore, this procedure can be extended to faulty oracles\footnote{Notice that a faulty oracle is not the same as an unreliable oracle, as in \cite{10.1145/226643.226647}}, faulty oracles of faulty oracles, and so on. This argument is quite similar to the argument which intuitively constructs the arithmetical hierarchy in computability, precisely because a process crash is a generalization of a non-halting computation.

Formalizing the previous argument, let $G = \langle V, E \rangle$ be a distributed system such that $u \in V$ is faulty. Then one constructs another $G' = \langle V \cup \{O_u\}, E \cup \bigcup_{i \in V} \{(i,O_u),(O_u,i)\}$, such that $O_u$ is a faulty detection oracle for $u$, and modifies the consensus algorithm $\mathcal{A}$ into a new algorithm $\mathcal{A}$' to include fault detection checks, as messages to and from $O_u$. Then:

\begin{enumerate}
    \item The phase transition in $G$ from $\mathcal{F}_c(C') = T$ to $\mathcal{F}_c(C') = F$, where $C'$ is a halting configuration, occurs when $g_\infty(G) = 2$.
    \item However, the same phase transition occurs in $G'$ when $g_\infty(G') = 2$.
\end{enumerate}

If a distributed system $G$ has a subset $P \subset V$ which is known to be faulty and $|P| = k$, then one can build a system $G^k$ with $k$ oracles, which then only undergoes a phase transition when $g_\infty(G) = k + 1$. Note that adding an oracle to $G$ increases the number of processes in the system and allows the oracle itself to be faulty, so $g_\infty$ does not differentiate between `types' of processes, that is, whether they are oracular or not. It is possible to consider a hierarchy of features, each of which measures the number of faulty processes as follows: $\mathcal{F}_0$ is a feature measured by the number of faulty processes $g_0$ that are not oracles, and $\mathcal{F}_n$ is a feature measured by the number of faulty oracles $g_n$ at level $n$ of the hierarchy. For example, $\mathcal{F}_2$ is measured by $g_2 : \mathbb{G} \rightarrow \mathbb{N}$ calculated as the number of faulty oracles of faulty oracles. Under this formalization, the same phase transition concerning consistency occurs when:

$$ \exists n~ \forall k~ (k > n \land g_k(G) = 0 \land g_n(G) = 1) $$

That is, there is at least one faulty oracle at the \textit{maximal} level of the oracular hierarchy in a given system $G$. Note that it is always possible to delay or postpone the phase transition of a system $G$ such that $g_n(G) = 1$ by adding an oracle and thus constructing a $G'$ such that $g_{n+1}(G') = 0$. But the phase transition is guaranteed to occur in $G'$ if it turns out that $G'$ has at least one faulty process at level $n+1$, albeit an oracular faulty one. Note that this assertion is precisely what characterizes emergent behavior according to the definition \ref{def:emergence}. Hence, one may argue that, even under the first perspective, inconsistency can be seen as emergent behavior under a more generalized definition of computation.

\subsection{Emergence and Non-Triviality}

Turning to logics, inconsistency as a property of a logic is entailed by the underlying axiomatic theory and its deduction rules. Stripping out a logic of all its deduction rules or its entire axiomatic theory guarantees consistency, but severely limits its deductive closure. Unlike distributed systems, inconsistency arises as a feature of the deductive machinery of a logic itself, and not as part of some piece or component of the logic, since one may argue that the deductive closure of a logic exists `in principle'.

Logics are not only concerned with their theoretical deductive closure, but also with the more general deductive closure over a set of premises. That is, given a logic $\mathbf{L}$, its theoretical deductive closure\footnote{In general, the deductive closure of a logic is denoted as $Cn(\cdot)$. The symbol $\mathcal{D}$ was chosen to reduce confusion with the logics $C_n$.} $\mathcal{D}_\mathbf{L}(\emptyset)$ may be inconsistent while its generalized deductive closure $\mathcal{D}_\mathbf{L}(\Gamma)$ is inconsistent, where $\Gamma$ is a set of formulas in the language $L$ of $\mathbf{L}$. If $\mathcal{D}(\Gamma) = L$ for every $\Gamma$, then $\mathbf{L}$ is trivial given $\Gamma$. Therefore, one may argue that, `in principle', a wide range of logics is \textit{inconsistent given a specific} $\Gamma$ --- for instance, take classical logic with $\Gamma_{\text{CPL}} = \{A, \lnot A\}$. But the same logic may not be \textit{trivial given the same} $\Gamma$ --- classical logic, in particular, trivializes given $\Gamma_{\text{CPL}}$ but the paraconsistent logic $C_1$ does not. In fact, the hierarchy of logics $C_n$ introduced by da Costa precisely captures this notion of constructing different logics each of which is able to \textit{delay} or postpone when trivialization of the deductive closure occurs.

Therefore, inconsistency as a feature of the deductive closure of a logic does not arise as emergent behavior, since one can always choose a set $\Gamma$ which itself is inconsistent. However, triviality may arguably be considered emergent behavior, specifically within the context of paraconsistent logics. Intuitively, the logic $C_n$ in the da Costa hierarchy trivializes its deductive closure for a set $\Gamma_n$, however, the same set $\Gamma_n$ does not trivialize the logic $C_{n+1}$. This may be seen as a phase transition, and, since this can be carried out indefinitely, triviality may be seen as emergent behavior. 

Let $\mathcal{F}_i$ and $\mathcal{F}_j$ denote the features of inconsistency and triviality of the deductive closure of a logic $\mathbf{L}$, measured by functions $g_i, g_j : \mathcal{P}(\mathbb{L}) \rightarrow \{T, F\}$, where $\mathcal{P}(\mathbb{L})$ denotes all subsets of all enumerable languages $\mathbb{L}$ of logics. Let $\Gamma \subset L$ be a set of formulas, then:

\begin{enumerate}
    \item If $g_i(\mathcal{D}_\mathbf{L}(\Gamma)) = T$, then certainly $\mathcal{D}(\Gamma)$ is inconsistent. One cannot infer from this that $g_i(\mathcal{D}_\mathbf{L}(\Gamma)) = T$, however the converse is \textit{true}.
    \item In particular, let $\Gamma_n$ be such that $g_j(\mathcal{D}_{C_n}(\Gamma_n)) = T$. Then, it is known that $g_j(\mathcal{D}_{C_{n+1}}(\Gamma_n)) = F$. But, by the definition of the da Costa hierarchy, one can construct a particular formula $A$ such that $g_j(\mathcal{D}_{C_{n+1}}(\Gamma_n \cup \{A\})) = T$.
\end{enumerate}

By the second point, one may argue that, in the da Costa hierarchy class of logics, inconsistency is not emergent behavior but triviality is: triviality appears through a phase transition and, although one can `delay' triviality in the deductive closure by moving up in the hierarchy, it is guaranteed to arise if one stops at a certain level. The argument here is very much the same as the one presented in the previous section.

Logics of formal inconsistency (LFIs) generalize the da Costa hierarchy by introducing a connective $\circ$ which acts as the consistency operator, capable of encoding when triviality occurs in the deductive closure of the logic as part of the language of the logic itself. That is, if $A \land \lnot A \land \circ A \in \mathcal{D}_\mathbf{L}(\Gamma)$, where $\mathbf{L}$ is an LFI, $\Gamma$ trivializes the deductive closure of $\mathbf{L}$, i.e. $\mathcal{D}_\mathbf{L}(\Gamma) = L$.

Suppose that $\{A, \lnot A, \circ A\} \in \mathcal{D}_\mathbf{L}(\Gamma)$ for an LFI $\mathbf{L}$, that is, $\Gamma$ trivializes $\mathbf{L}$. One may ask - is it possible to construct another LFI $\mathbf{L}'$ such that $\mathcal{D}_\mathbf{L'}(\Gamma)$ does not trivialize? Suppose so, then there is at least a formula $B \not \in \mathcal{D}_\mathbf{L'}(\Gamma)$. Notice that $\mathcal{D}_{\mathbf{L}}(\Gamma \cup \{B, \lnot B, \circ B\}) \neq L'$, where $L'$ is the language of $\mathbf{L}'$, as it would contradict the premise that $\Gamma$ does not trivialize $\mathbf{L'}$. Therefore, LFIs are able to bound inconsistency and triviality, at least when these properties arise out of a specific kind of structure, that is, out of contradictions. Then it may be argued that LFIs \textit{circumvent} triviality as emergent behavior. Although it is true that a logic such as $\mathbf{L}'$ may exist, one cannot assert that it is an LFI for certain.

\section{Conclusion}

This paper briefly presents how concepts from distributed computing, logic and complex systems interact. Taking oracles as generalized computation mechanisms, the consensus problem maintains its characteristics in the presence of a faulty process --- namely, one cannot attain termination and consistency simultaneously. However, in the realm of logic, inconsistency, as a property, has historically been shunned until the development of paraconsistency. Paraconsistency tames inconsistency and triviality, and, as this paper argues, certain classes of logic can circumvent triviality as a whole.

Although generalizations of the consensus problem are not a novel idea, the literature on distributed systems has devoted particular attention to the concept of oracles due to issues arising in the context of from blockchains and smart contracts \cite{CASINO201955, kustov2022mutual}. However, this paper presents another perspective on oracles for distributed systems as generalized computation mechanisms. In doing so, even in the presence of arbitrary computational power, the FLP impossibility result persists in the presence of at least one faulty process in a system. Due to this, what one this paper attempts to argue can be summarized in the following lemma:

\begin{quote}
    Inconsistency emerges from abstract computational consensus.
\end{quote}

As such, if one cannot ever attain consistency alongside termination, then it may be plausible to leverage paraconsistent logics as a reasoning tool for distributed systems. Although the consensus problem enjoys several solutions, perhaps the most well known and widely celebrated being the Paxos algorithm family due to Lamport \cite{10.1145/279227.279229, lamport2001paxos}, it is not uncommon to find distributed systems which, in practice, do not explicitly implement any particular solution. Even more so, the Paxos family of algorithms ensures consistency at the stake of having minimal probability of nontermination.

Paraconsistent consensus algorithms, or, consensus algorithms which implement paraconsistent logical reasoning, may be a good compromise in contexts where one cannot implement a well-known consensus algorithm, or where one needs to ensure theoretical termination, such as in real time systems. Even in domains in which inconsistencies are not acceptable, techniques such as repair semantics \cite{10.1145/3597304} could be an interesting venue for research to cope with these requirements.

\printbibliography

\end{document}